\documentclass[aps,longbibliography,showpacs,twocolumn,superscriptaddress,amsmath,amssymb,verbatim]{revtex4-1}
\usepackage{mhchem}
\usepackage[thinlines]{easytable}
\usepackage{graphicx}
\usepackage{lmodern}
\usepackage{epstopdf}
\usepackage{dcolumn}
\usepackage{bm}
\usepackage{subfigure}
\usepackage{makecell}
\usepackage{amsmath} 
\usepackage[pdftex,colorlinks=true,citecolor=blue,linkcolor=blue,urlcolor=blue,bookmarks=true]{hyperref}
\usepackage{booktabs}
\usepackage{multirow} 
\usepackage{array} 

\usepackage{color}
\usepackage{textcomp}
\definecolor{red}{rgb}{1,0,0}

\definecolor{blue}{rgb}{0,0,1}

\definecolor{green}{rgb}{0,1,0}

\makeatletter

\ProvideTextCommand{\~}{LGR}[1]{\char126#1}

\@ifundefined{textcolor}{}
{%
	\definecolor{BLACK}{gray}{0}
	\definecolor{WHITE}{gray}{1}
	\definecolor{RED}{rgb}{1,0,0}
	\definecolor{GREEN}{rgb}{0,1,0}
	\definecolor{BLUE}{rgb}{0,0,1}
	\definecolor{CYAN}{cmyk}{1,0,0,0}
	\definecolor{MAGENTA}{cmyk}{0,1,0,0}
	\definecolor{YELLOW}{cmyk}{0,0,1,0}
}

\makeatother

\makeatother

\begin{document}

\title{Exotic magnetism and persistent short-range  spin correlations in a frustrated honeycomb antiferromagnet}

\author{M. Barik}
\affiliation{Department of Physics, Indian Institute of Technology Madras, Chennai 600036, India}

\author{Q.~Faure}
\affiliation{Laboratoire Léon Brillouin, CEA, CNRS, Université Paris-Saclay, CEA-Saclay, 91191 Gif-sur-Yvette, France}

\author{F.~Damay}
\affiliation{Laboratoire Léon Brillouin, CEA, CNRS, Université Paris-Saclay, CEA-Saclay, 91191 Gif-sur-Yvette, France}

\author{J. P. Embs}
\affiliation{Laboratory for Neutron Scattering and Imaging, Paul Scherrer Institute, 5232 Villigen PSI,
Switzerland
}

\author{S.~Petit}
\affiliation{Laboratoire Léon Brillouin, CEA, CNRS, Université Paris-Saclay, CEA-Saclay, 91191 Gif-sur-Yvette, France}

\author{P. Khuntia}
\email{pkhuntia@iitm.ac.in}
\affiliation{Department of Physics, Indian Institute of Technology Madras, Chennai 600036, India}
\affiliation{Quantum Centre of Excellence for Diamond and Emergent Materials,
Indian Institute of Technology Madras, Chennai 600036, India}


\begin{abstract}
	Two-dimensional high-spin bipartite honeycomb networks, where anisotropy, competing exchange interactions, and spin fluctuations are at play offer an alternative route to test theoretical models to distinguish between classical and quantum magnetism in the context of emergent many-body phenomena with exotic excitations. Here, we present the crystal structure, magnetization, specific heat and inelastic neutron scattering experiments of an $S=5/2$ distorted honeycomb magnet CaZn$_2$Fe(PO$_4$)$_3$. The magnetization measurements reveal a dominant antiferromagnetic interaction between the Fe$^{3+}$ ($S=5/2$) moments. The development and field evolution of a dip under external magnetic field in the susceptibility suggests the emergence of unconventional field-induced transition, which is further supported by the anomalies observed in magnetization isotherms. The zero-field specific heat measurement reveals an antiferromagnetic transition at $T_\text{N}  \sim$ 1.67 K, that evolves with external magnetic field, suggesting the stabilization of a field-induced spin-canted state. Thermodynamic experiments reveal the presence of short-range spin correlations above the transition temperature. Inelastic neutron scattering results further corroborate an antiferromagnetic ordering, in agreement with the specific heat. Spin wave calculations reveal competing exchange interactions that induce magnetic frustration along with weak Ising-like anisotropy.  Competing exchange interactions and anisotropy lead to exotic field-induced phenomena and drive the system into close proximity of a mean field tricritical point in the $J_2/J_1-J_3/J_1$ phase diagram, opening an avenue to exotic states in high-spin frustrated honeycomb magnets.

\end{abstract}

\date{\today}

\maketitle

A synergistic interplay between emergent degrees of freedom, underlying symmetry, and competing exchange interactions in frustrated quantum magnets offers a promising platform for the experimental realization of exotic quantum states, including quantum spin liquids and field-induced many-body phenomena in contemporary condensed matter \cite{balents2010spin,khatua2023experimental,khuntia2020gapless,PhysRevLett.116.107203,PhysRevB.93.140408,PhysRevB.106.104404}. While strong quantum fluctuations in frustrated magnets can stabilize the celebrated QSL state, fluctuations in the critical limit can drive the system into a regime \citep{vojta2018frustration,PhysRevLett.119.165701,blanc2018quantum}, in which distinct ground states can be tuned by external perturbations. The application of a magnetic field can modify the singlet-triplet gap \citep{wang2006field}, while hydrostatic pressure can alter bond-lengths, effectively tuning the strength of exchange interactions \citep{PhysRevLett.93.257201}. Moreover, chemical substitution \citep{simonet2006effect}, which changes the orbital overlap, can enrich the phase diagram \citep{vojta2018frustration}, which is often characterized by unconventional critical exponents of order parameter correlations, dimensional crossover, and non-trivial symmetry breaking \citep{vojta2018frustration}. 

When the spin Hamiltonian is {effectively }invariant under \textit{U}(1) symmetry, reflecting the {approximate }conservation of boson number, a field-induced phase transition can lead to spontaneous breaking of the \textit{U}(1) symmetry, resulting in a Bose-Einstein condensation (BEC) of magnons in low dimensional magnets \citep{RevModPhys.86.563}, as observed in 1D chain CsFeBr$_3$ \citep{PhysRevB.96.144404}. In general, the presence of strong anisotropy, spin-orbit coupling, or Dzyaloshinskii-Moriya (DM) interactions in high-spin systems explicitly breaks the \textit{U}(1) symmetry \citep{PhysRevB.104.245107,giamarchi2008bose}, thereby preventing the system from undergoing BEC of magnons. Nevertheless, the presence of multiple excitation levels in such systems leads to complex field-induced phenomena including multiple field-induced BEC-like states \citep{watanabe2023double}, and spin supersolid phases \citep{PhysRevLett.98.227201}, thus expanding the magnetic phase diagram consisting of competing phases.

\begin{figure*}
    \centering
    \includegraphics[width=1\linewidth]{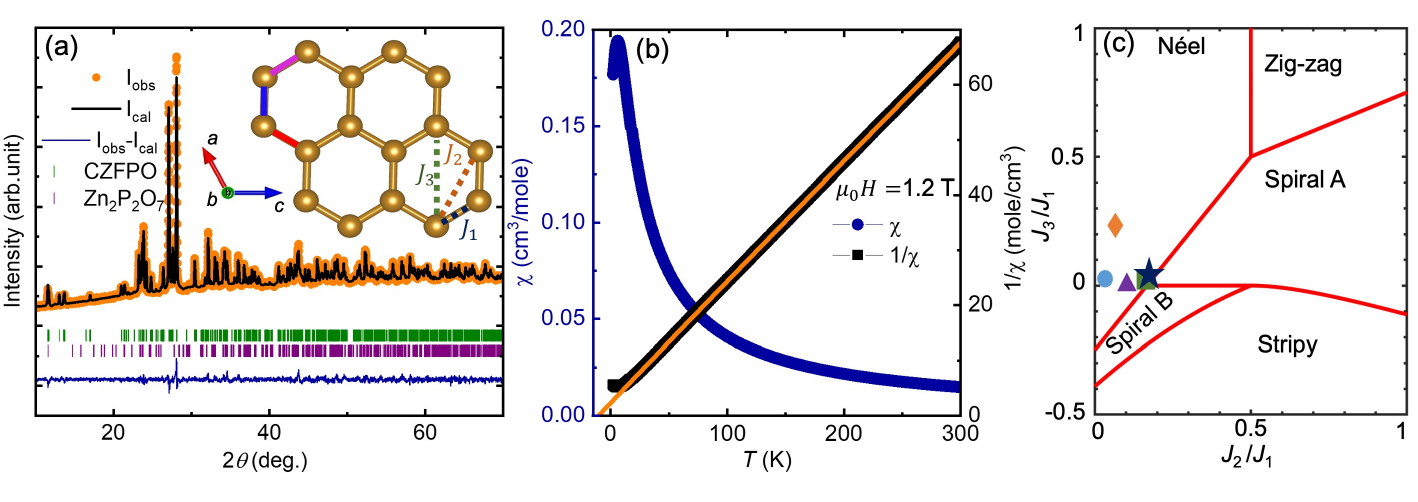}
    \caption{(a) Two-phase Rietveld refinement of the XRD pattern of polycrystalline CZFPO recorded at room temperature. The refinement confirms CZFPO as the dominant phase ($>99\%$) with a minor ($\approx0.1\%$) Zn$_2$P$_2$O$_7$ impurity identified by a peak at $29.5^\circ$. The inset illustrates the distorted honeycomb spin lattice with pink, blue and red bonds representing bond lengths of  4.761 \AA, 5.087 \AA ~and 4.790 \AA, respectively. As the bond lengths are comparable, nearest neighbor, second nearest neighbor and third nearest neighbors for each site are considered identical and represented as $J_1,J_\text{2}$ and $J_\text{3}$. (b) Temperature dependence of the dc magnetic susceptibility measured at 1.2 T. The solid orange line represents the Curie-Weiss fit to the inverse susceptibility in the paramagnetic region. (c) $J_3/J_1$ vs. $J_2/J_1$ phase diagram for a honeycomb lattice. The phase boundaries are plotted with reference to \cite{mazin2013camn,PhysRevB.91.180407}. The dark navy star denotes the position of the CZFPO in the phase diagram. The green square, purple triangle, sky blue circle and orange diamond are representation of positions of honeycomb antiferromagnets CaMn$_2$Sb$_2$ \cite{PhysRevB.91.180407}, MgMnO$_3$ \cite{PhysRevMaterials.3.124406}, VPS$_3$ \cite{PhysRevB.111.134421} and MnPS$_3$ \cite{wildes1998spin}, respectively.}
    \label{fig:1}
\end{figure*}

While honeycomb magnets hosting spin–orbit–driven bond-dependent anisotropic exchange interactions with $J_\text{eff}=1/2$ have been widely explored in the quest for the Kitaev QSL \cite{PhysRevB.102.014427} and unconventional field-induced phenomena such as 2D BEC \cite{matsumoto2024quantum}, high spin magnets on the honeycomb lattice offer a platform for diverse magnetic phases driven by magnetic anisotropy, competing exchange interactions and low coordination number. {In particular, when the competing exchange strengths $J_1$, $J_2$, $J_3$   representing the first, second and third nearest neighbor in-plane interactions, approach critical ratios, various competing magnetic phases can be tuned across the phase boundary by introducing external perturbations such as magnetic field and pressure \cite{fouet2001investigation}, thereby prompting for the experimental realization of exotic states in honeycomb antiferromagnets proximate to $J_2/J_1-J_3/J_1$ tricritical points, where magnetic frustration is maximized \cite{PhysRevB.91.180407,PhysRevMaterials.3.124406,mazin2013camn} (see Fig. \ref{fig:1}(c)). In such a near-degenerate regime, anisotropy can lift the degeneracy and select among competing magnetic states; for instance the single ion anisotropy stabilizes the honeycomb antiferromagnet Ba$_2$NiTeO$_6$ in a stripe ordered state \citep{PhysRevB.96.104414}.} Notably, such anisotropy can originate from structural distortions that break inversion symmetry, giving rise to DM interactions. These structural distortions can further influence the magnetic ground state by modifying the local crystal electric field (CEF) environment, bond angles and exchange interaction pathways, thereby enabling anisotropic exchange interactions. Such effects can stabilize canted spin configurations, drive spin reorientation \citep{dergachev2007spin,PhysRevB.72.174429}, and induce dimensional crossover and exotic phases in distorted honeycomb antiferromagnets \citep{wildes2006static}. {In this vein, the discovery and investigation of high-spin distorted honeycomb magnets, with competing exchange interactions, near-critical exchange ratios and magnetic anisotropy has the potential to host perturbation-induced exotic ground states, is highly relevant.}

Herein, we present the magnetic properties of an $S=5/2$ magnet, namely CaZn$_2$Fe(PO$_4$)$_3$, henceforth CZFPO, where Fe$^{3+}~(S=5/2)$ ions are arranged on a distorted honeycomb lattice in the \textit{ac}-plane. The magnetic interactions are predominantly antiferromagnetic,  as evidenced by the Curie-Weiss temperature of $-10.3(1)$ K. The system undergoes an antiferromagnetic N\'eel ordering at $T_\text{N}\sim$1.67 K, confirmed from the thermodynamic and inelastic neutron scattering (INS) experiments. Thermodynamic experiments  reveal robust short-range spin correlations persisting well above the transition temperature characteristics feature  of low dimensional magnets. {The well reproducibility of the specific heat for $T\ll T_\text{N}$ with 2D gapped excitation model consistent with the presence of anisotropy, observed from the spin wave calculations, which yields exchange interactions $J_1=2.08$~K, $J_2= 0.348$~K, $J_3=0.023$~K and $D=0.069~$K.} The system undergoes field-induced spin canted state with the application of magnetic field, as revealed by the field evolution of minimum in the magnetic susceptibility and $T_\text{N}$ in the specific heat. The intriguing field-induced phenomenon in this frustrated $S = 5/2$ honeycomb magnet arises from the interplay between competing exchange interactions, which stabilize a nearly degenerate manifold of magnetic ground states, and weak anisotropy in the system. CZFPO is located in the vicinity of a tricritical point, which most likely enhances frustration induced fluctuations, leading to a reduced $T_\text{N}$.

The title compound, CZFPO, was synthesized following a standard solid-state reaction route, as described in the SI \cite{SM}. The material contains an unavoidable minor nonmagnetic impurity phase, Zn$_2$P$_2$O$_7$ ( $\sim$ 0.1\%), which is frequently observed in polycrystalline samples \cite{arh2022ising,PhysRevB.108.054442}. Fig. \ref{fig:1}(a) depicts the two-phase Rietveld refinement of CZFPO, with the refinement parameters provided in the Ref. \cite{SM}. CZFPO crystallizes in a monoclinic crystal structure with the space group $\textit{P}2_1$/c (see Ref. \cite{SM}). The Fe$^{3+}$ ions constitute a distorted honeycomb spin-lattice perpendicular to the \textit{b}-axis as shown in the inset of Fig. \ref{fig:1}(a). The Fe$^{3+}$ ions reside in an octahedral environment connected to PO$_4$ tetrahedra, which mediates the exchange interaction via the Fe-O-P-O-Fe superexchange pathway.
\begin{figure*}
    \centering
    \includegraphics[width=1\linewidth]{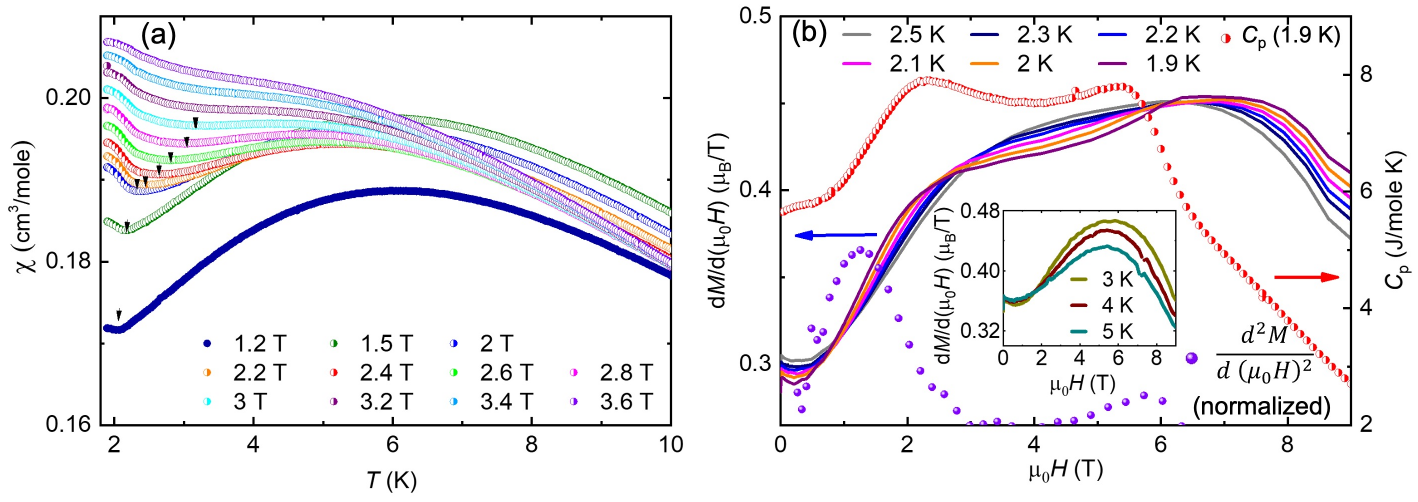}
    \caption{(a) Temperature dependence of the magnetic susceptibility measured at various magnetic fields. The black arrows indicate the minimum. The transition temperature is calculated from the minimum in $d\chi/dT$ as shown in the Ref. \cite{SM}. (b) Field dependence of $dM/d(\mu_0H)$ (left panel) and $C_\text{p}$ (right panel). The violet spheres represent the normalized $d^2M/d(\mu_0H)^2$ at 1.9 K. The inset shows $dM/d(\mu_0H)$ for 3 K $\leq T \leq$ 5 K.}
    \label{fig:2}
\end{figure*}

The magnetic susceptibility recorded at 1.2 T in the range 1.9 K $\leq$ \textit{T} $\leq$ 300 K is depicted in the Fig. \ref{fig:1}(b), where a broad maximum around 6 K, indicates short-range spin correlations. The Curie-Weiss fit to the inverse susceptibility at 1.2 T in the range 200 K $\leq T\leq$ 300 K, yields a Curie-Weiss temperature ($\theta_\textnormal{CW})\approx -$10.3(1) K, revealing a dominant antiferromagnetic exchange interaction between the Fe$^{3+}$ moments with an effective magnetic moment
 $\mu_\textnormal{eff}$ $\approx$ 5.99 $\mu_\textnormal{B}$, close to that expected for a Fe$^{3+}$ ion. The overlap of zero-field cooled (ZFC) and field cooled (FC) susceptibilities at an applied field of 100 Oe rules out the presence of frozen spins down to at least 5 K (see Ref. \cite{SM}). 

 With an increase of magnetic field ($\mu_0H\geq$ 1.2 T), the development of a minimum in the susceptibility, followed by its shift towards high temperature under the application of high magnetic field is observed (see Fig. \ref{fig:2}(a)).  This behavior resembles field-induced transitions in frustrated magnets CsFeBr$_3$ \citep{tanaka2001field}, and K$_2$Ni$_2$(MoO$_4$)$_3$ \citep{PhysRevB.95.180407} associated with BEC of exotic excitations, in contrast to the conventional LRO. The field-induced behavior is further corroborated by isothermal magnetization data taken at 1.9 K (see SI ref. \cite{SM}). {The field dependence of $dM/d(\mu_0H)$,  reveals a minimum  around 0.4 T, which has also been observed in several frustrated magnets \citep{PhysRevB.95.180407,watanabe2023double}, intensifies with decreasing temperature, most likely due to the realignment of the spins \citep{watanabe2023double} (see  Fig. \ref{fig:2}(b)). Upon decreasing the temperature, even while remaining above $T_\text{N}=1.67$~K (see the next section), the peaks in $dM/d(\mu_0H)$ at 2 T  and 6 T become more pronounced, indicating enhanced spin reorientations as the temperature approaches $T_\text{N}$ \citep{watanabe2023double}, rather than conventional spin-flop transition, which typically occurs below $T_\text{N}$ \citep{PhysRevB.110.184402}.} At high temperatures ($T\geq$ 3 K), $dM/d(\mu_0H)$ exhibits one single dome, in contrast to the two weak kinks observed below 2.5 K (see inset of Fig. \ref{fig:2}(b)), suggesting the emergence of unconventional spin reconfigurations below 2.5 K \citep{haseda1981spin}. To further investigate the field-induced phenomena, the specific heat measurements were performed at 1.9 K under magnetic field upto 9 T. At 1.9 K, both $dM/d(\mu_0H)$ and $C_\text{p}$ curves show an anomaly at around 2 T, corresponding to the lower critical field (see Fig. \ref{fig:2}(b)). In contrast to the sharper transition in  $C_\text{p}$($\mu_0H$) at $\mu_0H_{c2}$, $dM/d(\mu_0H)$ exhibits a distribution of critical field, possibly due to the competing field-induced magnetic states \citep{qureshi2021field,PhysRevB.91.014406,PhysRevLett.101.187205}. To estimate $\mu_0H_{c2}$, the peak in the second derivative of $M$ with respect to $\mu_0H$, which coincides with the $\mu_0H_{c2}$ obtained from $C_\text{p}$($\mu_0H$), has been taken into account \citep{qureshi2021field,PhysRevB.91.014406,PhysRevLett.101.187205}.

 \begin{figure*}
     \centering
     \includegraphics[width=1\linewidth]{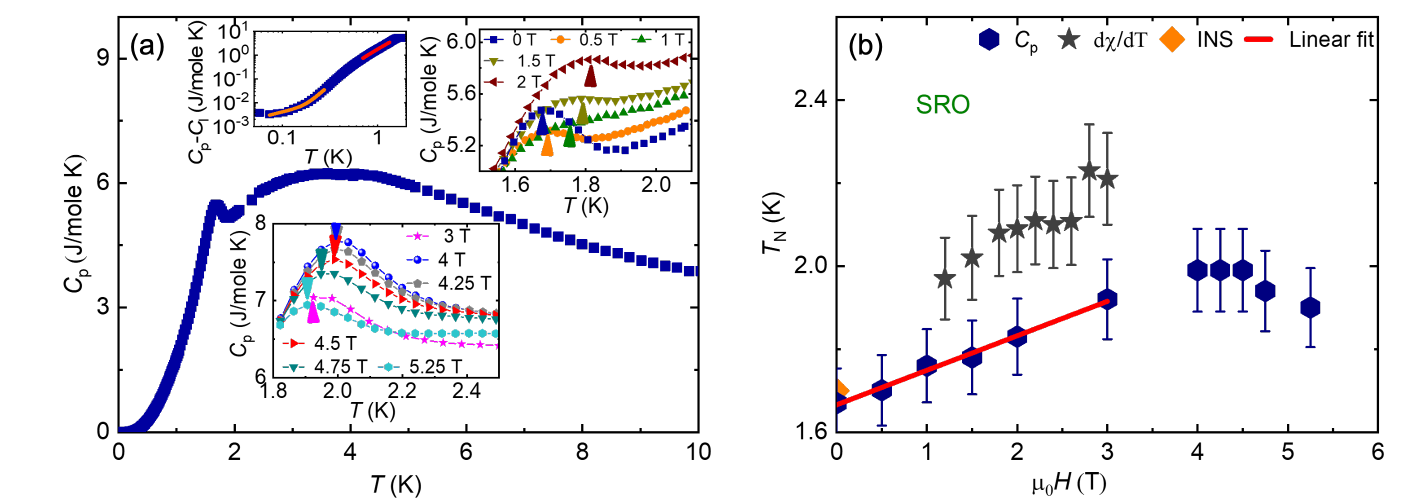}
     \caption{(a) Temperature dependence of the zero-field specific heat. Top left inset: Temperature dependence of zero field magnetic specific heat below 2 K. The orange curve represents a fit to Eq. \ref{equation1} for $T \leq 0.3$ K, while the red line highlights a $T^{2.17}$ power-law behavior. Top right inset: $C_\text{p}$ illustrating an increase of $T_\text{N}$ with magnetic field. Bottom inset: $C_\text{p}$ at higher magnetic fields where $T_\text{N}$ decreases with field. The arrows in the insets indicate the transition temperatures. To identify the onset of magnetic ordering, we have presented the corresponding inflection points in $dC_\text{p}/dT$ in Ref.~\cite{SM}. (b) Magnetic phase diagram of CZFPO constructed from $T_\text{N}$ obtained across various experiments. Vertical bars denote error bars. Red solid line represents the linear fit to the transition temperatures at low field. }
     \label{fig:3}
 \end{figure*}

\begin{figure*}
    \centering
    \includegraphics[width=\linewidth]{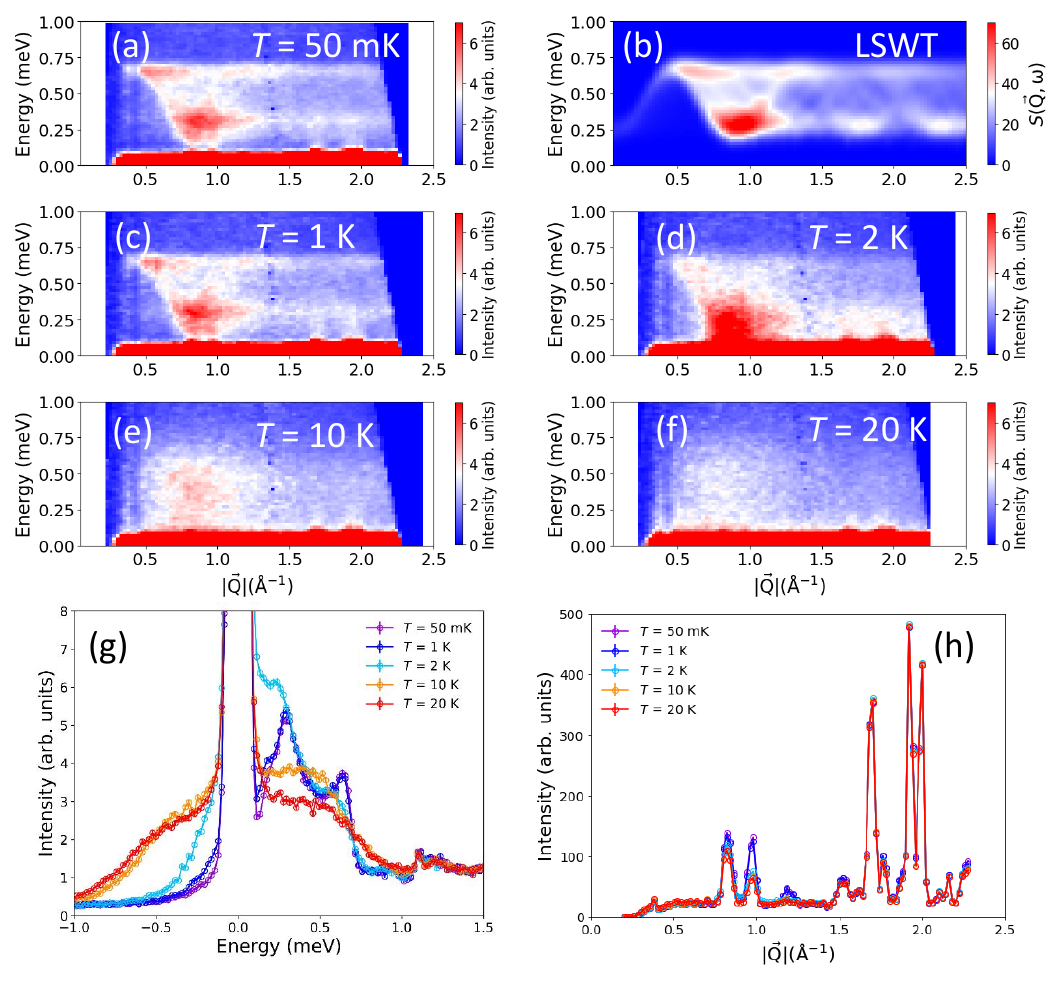}
    \caption{Inelastic neutron scattering spectra. (a) Experimental INS spectrum collected at $T = 50$~mK. (b) Calculated spectrum based on linear spin wave theory (LSWT) using $J_1 = 2.08$~K, $J_2 = 0.35$~K, $J_3 = 0.023$~K and $D = 0.069$~K. (c-f) Experimental INS spectrum at $T = 1, 2, 10$ and $20$~K, respectively. (g) Temperature evolution of the integrated intensity over $\boldsymbol{Q}$ range $[0.7-1.2]~\AA^{-1}$. (h) Temperature evolution of the elastic line, obtained by integrating data over the energy range $[-0.1-0.1]$~meV.}
    \label{INS-mesure}
\end{figure*}
To shed further insights into the magnetic phases, zero field and field dependent specific heat measurements (see Fig. \ref{fig:3} (a) and inset), were performed.  The zero field specific heat exhibits a broad maximum around 4 K, indicative of short-range spin correlations, consistent with the magnetic susceptibility. A sharp anomaly at $T_\mathrm{N} = 1.67$ K signals the onset of long-range magnetic order. {After subtracting the lattice  specific heat (see Ref \cite{SM}), at low temperatures ($T \leq 0.3$ K),  the specific heat follows asymptotic 2D gapped excitation model approximating the energy as $\varepsilon_{\mathbf{k}} \approx \Delta + c k^2$ \cite{PhysRevB.103.L180406}, given by,
\begin{multline}
   C_\text{p}-C_\text{l}  \\=T \left[ \frac{1}{c}\left(\left(\frac{\Delta}{k_\text{B}T}\right)^2 + 2\left(\frac{\Delta}{k_\text{B}T}\right) + 2\right)e^{-(\frac{\Delta}{k_\text{B}T})} + b  \right] 
   \label{equation1}
\end{multline}
where $c$ is the bandwidth parameter and $b$ is the offset. The obtained gap $\Delta/k_\text{B}= 1.38$ K  is consistent with a gap revealed in the INS spectrum, originating from single-ion anisotropy (see next section). Equation (\ref{equation1}) applies only in the low-temperature limit $T \ll \Delta/k_\text{B}$ near the dispersion minimum \cite{PhysRevB.103.L180406}. In the range, $0.7 ~\text{K} \leq T \leq 1.3 $ K, the specific heat undergoes a crossover to power-law behavior $T^{2.17}$, which further confirms the 2D correlations among the excitations below $T_\text{N}$ \cite{PhysRevB.96.041405,PhysRevLett.99.097201}.} With increase of magnetic field, $T_\textnormal{N}$ shifts towards high temperature up to 4 T, in contrast to conventional antiferromagnets, where the $T_\textnormal{N}$ typically decreases with increasing of the field, { indicating a field-driven gradual tilting of the spins away from the antiferromagnetic configuration \citep{PhysRevX.9.011038,tanaka2001field, PhysRevB.96.144404}. Upon further increasing the field, as the spins approach full polarization, the transition temperature decreases (see bottom inset of Fig. \ref{fig:3}(a) ). This shift in $T_\text{N}$ with applied field ($\mu_0H$) is indicative of a gradual reorientation of the spin structure from a collinear antiferromagnetic state towards a canted spin configuration \citep{PhysRevX.9.011038}. Notably, CZFPO retains short-range spin correlations even under high field, signalling the robustness of magnetic frustration in this system (see Ref \cite{SM}).

Fig. \ref{fig:3}(b) depicts the evolution of the $T_\textnormal{N}$ with magnetic field extracted from specific heat, consistent with magnetic susceptibility measurements. The difference in $T_\text{N}$ in both of these techniques arises from the distinct underlying measurement principles inherent to each method \citep{PhysRevB.104.L060411,PhysRevB.109.054405}. The $T_\textnormal{N}$ vs. $\mu_0H$ replicates a dome-like behavior, as observed in several frustrated magnets, suggesting the realization of an unconventional field-induced phase \citep{PhysRevB.95.180407,PhysRevX.9.011038,watanabe2023double}. In the present magnet, the behavior may be ascribed to field-induced spin canting mediated by anisotropic interactions, which originates from the combined effects of finite single-ion anisotropy of Fe$^{3+}$ ions and inversion symmetry breaking along the Fe$^{3+}-$Fe$^{3+}$ bond, due to the structural distortion in the honeycomb plane. A similar evolution of $T_\text{N}$ has been observed in Ba$_2$FeSi$_2$O$_7$ \citep{PhysRevB.104.214434} and Ba$_3$Mn$_2$O$_8$ \citep{PhysRevB.77.214441} in their canted phases. {The nature of the field-induced state can be determined from the universality class governing the phase boundary \citep{giamarchi2008bose}. The observed linear dependence of transition temperature in low fields deviates from the characteristic power law behavior ($T_\text{N}\propto \mu_0(H-H_C)^{2/3}$) expected for the three dimensional BEC \citep{PhysRevB.95.180407}. Instead, such a quasi-linear ($T_\text{N}\propto \mu_0(H-H_C)$) phase boundary is more consistent with reduced dimensionality effects, as anticipated for quasi-two-dimensional BEC \citep{RevModPhys.86.563,PhysRevB.37.4936,giamarchi2008bose} or Bose glass scenario \citep{watanabe2023double,yu2012bose}. Given that the Bose-glass phase emerges in the presence of disorder, the phase boundary in the field-induced ordered state may represent the crossover from the short-range ordered state to quasi 2D-BEC state, as suggested by the quasi-2D nature of the excitations evidenced from specific heat experiments.} Such quasi 2D BEC state arises when interlayer coupling is weak or frustrated \cite{PhysRevB.77.094406}, leading to effectively two dimensional critical behavior \citep{RevModPhys.86.563,watanabe2023double,PhysRevB.77.094406,PhysRevB.37.4936,sebastian2006dimensional,matsumoto2024quantum,sheng2025bose}}.

Inelastic neutron scattering technique is an excellent probe for detecting competing ground states and associated low-energy excitations in frustrated honeycomb magnets~\cite{arh2022ising}. To shed insights into ground state, exchange  interaction, and spin correlations, we performed INS experiment down to 50 mK at the TOF spectrometer FOCUS installed at PSI (Switzerland). The elastic line, defined by integrating data over the energy range $[-0.1-0.1]$ meV, is particularly informative. As shown in Fig.~\ref{INS-mesure}(h), it displays at least three Bragg peaks, which appear below about 1.7~K, confirming a
$\boldsymbol{Q} = {0}$ magnetically ordered ground state, consistent with the thermodynamic results.

The dynamic structure factor $S(\bm{Q},\omega)$ maps at various temperatures are presented in Figure~\ref{INS-mesure}. The spectra consist of  a dispersing band below 0.65~meV, reaching a minimum at the positions of the Bragg peaks, i.e. around 0.8~\AA$^{-1}$, however a small gap of about 0.2~meV persists at the lowest temperature, most likely reflecting the presence of weak anisotropy in the underlying spin lattice. With increasing temperature, this spectrum progressively softens. Above the critical temperature, between 4.5~K up to 10~K, the data exhibit a broad feature centered at 0.8~\AA$^{-1}$, whose intensity decreases with increasing temperature  while no discernible signal is detected in the 20~K to  50~K temperature range. Energy cuts were also obtained by integrating data in $[0.7-1.2] $~\AA$^{-1}$, as shown in Figure~\ref{INS-mesure}(g).

{The interaction Hamiltonian of similar distorted honeycomb lattice are well approximated by the $J_1-J_2-J_3$ model in honeycomb magnets CaMn$_2$Sb$_2$ \cite{PhysRevB.91.180407}, NiPS$_3$ \cite{PhysRevB.98.134414} and Li$_2$IrO$_3$ \cite{PhysRevB.84.180407}. In CZFPO, the nearly equivalent bond lengths, a common exchange pathway, and comparable bond angles along the superexchange route support  modelling the spin Hamiltonian in the same framework ~\cite{fouet2001investigation, PhysRevB.84.094424} with single-ion anisotropy as a first approximation:}
\begin{multline}
\mathcal{H} = J_1\sum_{\langle i,j \rangle} \bm{S}_{i} \cdot \bm{S}_{j} + J_2\sum_{\langle \langle i,j \rangle \rangle} \bm{S}_{i} \cdot \bm{S}_{j} + J_3\sum_{\langle \langle \langle i,j \rangle \rangle \rangle} \bm{S}_{i} \cdot \bm{S}_{j} \\ - D \sum_{i} \bm({S}^{z}_{i})^2 
\label{Hamiltonian}
\end{multline}
where $J_1$, $J_2$, $J_3$ represent the first, second and third nearest neighbor in-plane exchange interactions, respectively, and $D$ denotes the single ion anisotropy; the later accounts for the observed spin gap of about 0.2 meV at 50~mK (Fig.~\ref{INS-mesure}(a)). Fig.~\ref{INS-mesure}(b) shows linear spin-wave calculations performed using the {\sc SpinWave} software ~\cite{Petit2010, Petit2011, Petit2016}. A N\'eel-type ground state with magnetic moments aligned along the $b$-axis, i.e. perpendicular to the honeycomb plane, was assumed, consistent with the three Bragg peaks and the $\bm{k} = \bm{0}$ magnetic structure observed in Fig.~\ref{INS-mesure}(h). The calculations were convoluted with the magnetic form factor of Fe$^{3+}$ magnetic ions. The best fit to the experimental data at 50 mK was obtained for $J_1$ = 2.08 K, $J_2$ = 0.35 K,  $J_3$ = 0.023 K, $D$ = 0.069 K. 

 Albeit a weak Ising anisotropy of $D = 0.069$~K, the experimental spin-dynamics is well captured by the frustrated $J_1-J_2-J_3$ model on a  distorted honeycomb lattice. In particular, the model reproduces the spin gap of about 0.2~meV and non-dispersive features around 0.3 and 0.6~meV above $|\bm{Q}|\simeq 1.5$~\AA$^{-1}$. Notably, the calculated spectra is consistent with a N\'eel ordered state with moments perpendicular to the honeycomb plane, i.e. with $D > 0$ (calculations with $D < 0$, (not shown) yield a qualitatively different spectrum). However, some discrepancies between experimental and calculated spectra remain, which may arise from (i) the distorted nature of the honeycomb lattice and/or (ii) interplane exchange interactions. Interestingly, extracted  exchange parameters locate CZFPO in close vicinity to a tri-critical point in the mean field $J_3/J_1-J_2/J_1$ phase diagram~\cite{PhysRevB.91.180407,fouet2001investigation,mazin2013camn,rastelli1979non} (see the  Fig. \ref{fig:1} (c)) and thus, external stimuli such as isostatic pressure or uniaxial strain could drive this system across phase transitions from a Néel phase to different spiral phases. For comparison, Fig. \ref{fig:1} (c) represents the $J_2/J_1$ and $J_3/J_1$ of honeycomb antiferromagnets CaMn$_2$Sb$_2$ \cite{PhysRevB.91.180407}, MgMnO$_3$ \cite{PhysRevMaterials.3.124406}, VPS$_3$ \cite{PhysRevB.111.134421} and MnPS$_3$ \cite{wildes1998spin}, all of which stabilize N\'eel ordered ground states. {Although the buckled honeycomb magnet CaMn$_2$Sb$_2$ also lies close to the tricritical point \cite{PhysRevB.91.180407}, CZFPO provides a better structural realization of a honeycomb lattice.} Within this model, the Curie Weiss temperature expressed as $\theta_{\rm{CW}} = -S(S+1)(J_1+2J_2+J_3)/k_{\rm{B}}$, yields $\theta_{\rm{CW}} \simeq -24.01$~K, slightly larger in magnitude than the experimental value determined from magnetic susceptibility measurements.
 
\setlength{\parskip}{0.5 em}
\par
To summarize, the distorted honeycomb magnet CZFPO undergoes a magnetic transition at 1.67 K, which is confirmed by zero field specific heat and INS measurements, despite the presence of competing exchange interactions in this frustrated magnets, in agreement with the calculated relative exchange interaction strengths. The spin-wave calculations within the $J_1$–$J_2$–$J_3$ model reveal weak Ising-like anisotropy and a gapped excitation spectrum, {with the latter independently corroborated by specific heat, drive the magnetism of the system}. Magnetic susceptibility and specific heat experiments reveal the presence of short range spin correlations well above $T_\text{N}$. The field evolution of transition temperature in the magnetization and specific heat measurements, is suggestive of an exotic field-induced state. The linear variation of the transition temperature at low fields indicates the field-induced canting is non-3D BEC scenario; instead, it indicates the 2D correlations of the condensates. The frustration induced by competing exchange interactions places the present material proximate to mean field tricritical point in the  $J_2/J_1-J_3/J_1$ phase diagram, thereby experimentally realizing a regime hitherto predicted theoretically. CZFPO thus emerges as a promising candidate
for realizing a rich tunable magnetic phase diagram controllable by
both magnetic field and external pressure due to its exchange
strengths with near critical exchange ratios. Future studies on high-quality single crystals may further reveal exotic competing phases driven by external stimuli.

\setlength{\parskip}{1 em}
\par 
\textit{Acknowledgements.} P.K. acknowledges the funding by the Anusandhan  National Research Foundation (ANRF), Department of Science and Technology, India through Research Grants. We also acknowledge PSI for neutron beam time and the 2FDN for financial support. 
\setlength{\parskip}{0.5 em}
\par
\textit{Data availability.}
The data that support the findings of the current study are available from the corresponding author upon reasonable request.


\bibliography{CZFP2}

\end{document}